\documentclass[]{interact}

\usepackage{epstopdf}  
\usepackage{booktabs}
\usepackage{rotating}
\usepackage{graphicx}
\usepackage{verbatim}  

\usepackage{amsmath}  
\usepackage{amssymb}  
\usepackage{hyperref}
\usepackage{cleveref} 
\usepackage{lmodern}  
\usepackage{multirow}
\usepackage{caption}
\usepackage{subcaption}
\usepackage{adjustbox}

\usepackage{natbib}
\theoremstyle{plain}  

\theoremstyle{definition}

\theoremstyle{remark}

\begin{document}


\title{Effects of Material Mapping Agnostic Partial Volume Correction for Subject Specific Finite Elements Simulations}

\author{
  \name{A. Beagley\textsuperscript{a}\thanks{CONTACT A. Beagley. Email: abeagley@uvic.ca}, H. Richards\textsuperscript{a}, and J. W. Giles\textsuperscript{a}}
  \affil{\textsuperscript{a}Orthopaedic Technologies and Biomechanics Lab, University of Victoria, Victoria, BC, Canada}
}

\maketitle

\begin{abstract}
  Partial Volume effects are present at the boundary between any two types of material in a CT image due to the scanner's Point Spread Function, finite voxel resolution, and importantly, the discrepancy in radiodensity between the two materials. In this study a new algorithm is developed and validated that builds on previously published work to enable the correction of partial volume effects at cortical bone boundaries. Unlike past methods, this algorithm does not require pre-processing or user input to achieve the correction, and the correction is applied directly onto a set of CT images, which enables it to be used in existing computational modelling workflows. The algorithm was validated by performing experimental three point bending tests on porcine fibulae specimen and comparing the experimental results to finite element results for models created using either the original, uncorrected CT images or the partial volume corrected images. Results demonstrated that the models created using the partial volume corrected images did improved the accuracy of the surface strain predictions. Given this initial validation, this algorithm is a viable method for overcoming the challenge of partial volume effects in CT images. Thus, future work should be undertaken to further validate the algorithm with human tissues and through coupling it with a range of different finite element creation workflows to verify that it is robust and agnostic to the chosen material mapping strategy.
\end{abstract}


\begin{keywords}  
  Partial Volume Artifacts; Finite Element; Computed Tomography
\end{keywords}

\section{Introduction}
To derive accurate subject-specific finite element (FE) models from patient data, two key considerations are accurately replicating skeletal geometry and material properties \citep{knowlesQuantitativeComputedTomography2016, eberleIndividualDensityElasticity2013,schileoAccurateEstimationBone2008,   yosibashPredictingYieldProximal2010a,keyakComparisonSituVitro2003a,    szwedowskiSensitivityAnalysisValidated2011,BabazadehNaseri2021,Vaananen2019,Taddei2007,pauchardInteractiveGraphcutSegmentation2016,Helgason2008,Helgason2016,Pakdel2016}. Geometric accuracy can be achieved using automatic meshing procedures \citep{Falcinelli2016} after segmenting computed tomography (CT) images \citep{pauchardInteractiveGraphcutSegmentation2016,luAccurate3DBone2016,phellanMedicalImageSegmentation2016}. Heterogenous mechanical properties can be obtained from the density calibrated Hounsfield units of the CT images using correlations with bone mineral density and empirically derived relationships between density and elastic modulus values \citep{BabazadehNaseri2021,Fung2017,Collins2021,Studders2020a}. These correlations and relationships depend on the anatomical site, density regime (cortical versus trabecular), and species \citep{liExperimentalNumericalStudy2020,fengMechanicalPropertiesPorcine2012,lesEstimationMaterialProperties1994,lindeMechanicalPropertiesTrabecular1991,snyderEstimationMechanicalProperties1991}. The development of these correlations and relationships is ongoing as multiple research groups have reported different results for the same anatomical site and species \citep{knowlesQuantitativeComputedTomography2016,Emerson2013,eberleInvestigationDetermineIf2013}.

Assigning these heterogenous material properties to FE meshes is non-trivial due to the difference in geometry and topology between the CT voxels and mesh elements. Various material mapping methods (MMMs) have been proposed that can primarily be categorized as either element-based or node-based. Element-based methods assign property values to each FE mesh element \citep{Taddei2007} while node-based methods assign property values to FE mesh nodes, allowing for material property variation across an element \citep{Helgason2008,BabazadehNaseri2021}.


Partial Volume (PV) effects are often observed in CT images and occur because each voxel represents the attenuation properties of the material(s) within the voxel's specific volume, meaning that if the volume contains multiple materials than the resulting Hounsfield Unit (HU) intensity represents some average of their properties \citep{Bushberg2020}. PV effects are most evident at boundaries between materials with markedly different radiodensities (e.g. air and cortical bone) causing sharp boundary to appear blurred. Additionally, all material boundaries in a CT image are blurred due to scanner system resolution limitations \citep{Pakdel2014,Bushberg2020}. PV effects have been shown to cause multiple challenges:

\begin{itemize}
    \item Incorrect diagnosis/assessment of cortical bone thickness \citep{Treece2015,Museyko2017}
    \item Artificially low cortical bone mineral density (BMD) \citep{Soucek2015}
    \item Increased difficulty in accurately segmenting structures \citep{Rittweger2004,Falcinelli2016,Peleg2014}.
\end{itemize}

In the context of deriving FE models from CT images, regardless of the MMM strategy used, surface nodes and elements may correspond to regions of the CT images that are affected by partial-volume (PV) effects. For skeletal FE models, PV effects typically result in underestimation of cortical bone density and therefore reduce the estimated elastic modulus \citep{Soucek2015,Peleg2014}, which can result in inaccurate FE simulation results.


Although various reconstruction kernels exist to minimize the effects of blurring caused by tomographic projection, they are not able to remove all blurring and PV effects in CT images. As a result, post-hoc methods have been developed to try to overcome the continued challenge of PV effects.
\cite{Helgason2008} introduced a modified MMM for FE meshes derived from CT scans that integrated a partial volume correction (PVC) method. After assigning material properties to mesh nodes, nearest neighbour interpolation was used to correct surface nodes that had a lower density than adjacent interior nodes \citep{Helgason2008}. \cite{Pakdel2016} built upon the previous method and introduced the Node-based elastic Modulus Assignment with Partial-volume correction (NMAP) method for material mapping of FE meshes derived from CT scans. This method was intended to be used after deblurring techniques are applied to reduce to effects of the CT scanner point spread function \citep{Pakdel2012,Pakdel2014,Pakdel2016}. NMAP performs the partial volume correction prior to mapping material properties onto the FE mesh by using an inverse distance weighted interpolation method to correct surface voxels in the segmentation of the CT scan itself.

Both methods demonstrated higher fidelity compared to experimental data than the widely used used material mapping methods available as part of the Bonemat software \citep{Taddei2007} and assigned material properties to mesh nodes instead of elements. However, neither method has an open-source implementation available for use. As well, the PVC methods are integrated with the material mapping strategies and have not been validated on their own, meaning they cannot be integrated into existing workflows to improve research results. For NMAP, Pakdel et al. notes that it has only been validated on CT scans that have undergone deblurring first, which is nontrivial to perform, making the method even more difficult to integrate into existing and varied workflows.

Therefore, the purpose of this study is to develop, validate, and release an open-source implementation of a PVC method that works on the CT scan itself and thus enabling either type of material mapping strategy to be used when deriving FE meshes from the CT scan. The specific goals in support of this purpose were to:

\begin{itemize}
    \item create a PVC method not depend on deblurring or other pre-processing techniques
    \item ensure the method can generalize to long bones, flat bones, etc.
    \item be simple to use (i.e. not require user expertise)
    \item make as few assumptions about the bone and partial volume effects as possible
    \item be deterministic given the same inputs
    \item have minimal inter-operator differences in results (e.g. not require thresholding to set the thickness of the PV layer as was done by Pakdel et al.)
\end{itemize}


\section{Materials and Methods}

\subsection{Experimental Model}
\begin{figure}[!htbp]
    \centering
    \includegraphics[width=\textwidth]{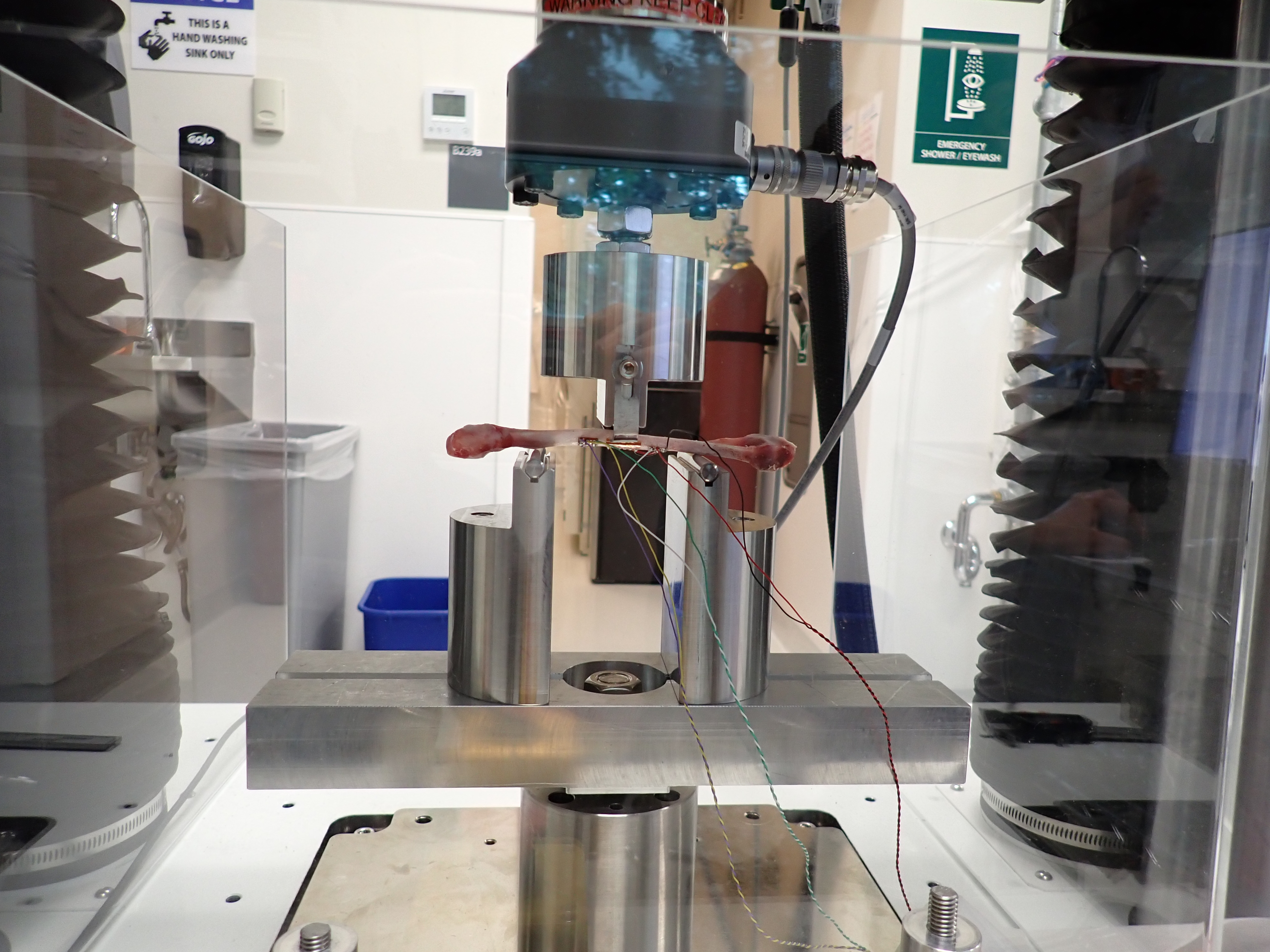}
    \caption{Experimental testing jig mounted to the MTS with a prepared fibula resting on the supports of the three-point bending jig. The affixed strain rosette can be seen on the fibula surface opposite the center driver, which is mounted to the load cell attached to the MTS mover.}%
    \label{fig:3bend_jig}
\end{figure}

Nine fresh frozen porcine hind limbs were obtained from a local abattoir. Ethics approval was not required for this study as the animal specimen were not sacrificed specifically for this study. The specimens were dissected to remove and denude the fibula and tibia. Three-point bend tests were performed for each fibula using a procedure adapted from the American Society of Agricultural and Biological Engineers (ASABE) Shear and Three-Point Bending Test of Animal Bone standard \citep{Asabe2005}. A custom bending jig (see \cref{fig:3bend_jig}) was designed and manufactured in accordance with \citep{Asabe2005} for use with an ElectroForce 3510 (TA Instruments, New Castle, Delaware) mechanical testing system to apply the bending load.

\begin{figure}[!htbp]
    \centering
    \begin{subfigure}[c]{0.48\textwidth}
        \centering
        \includegraphics[width=\textwidth]{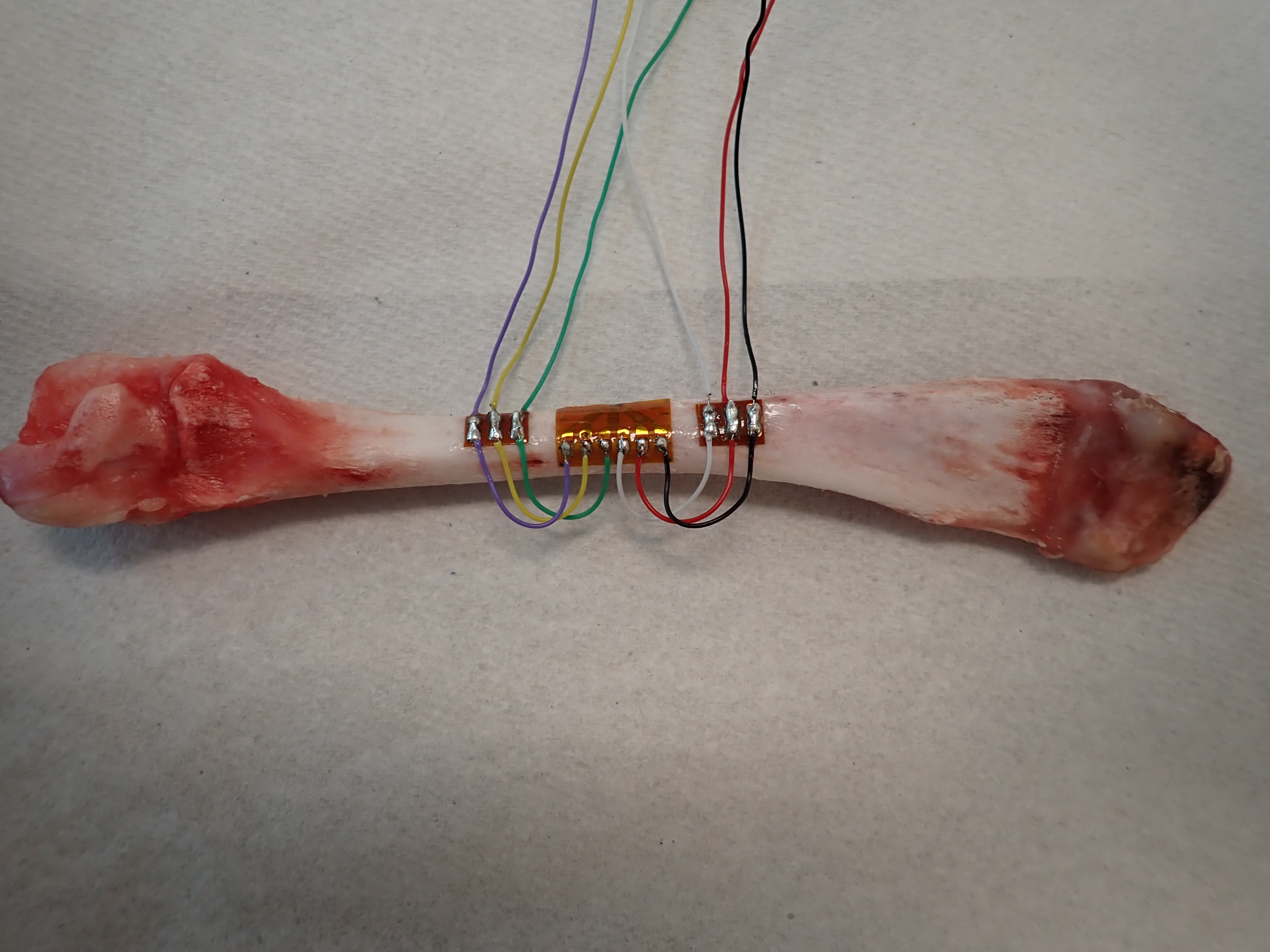}%
        \caption{Dissected fibula with a rectangular strain rosette affixed on the flattest surface at the center of the supported span.}
        \label{fig:fib_rosette}
    \end{subfigure}
    \hfill
    \begin{subfigure}[c]{0.48\textwidth}
        \centering
        \includegraphics[width=\textwidth]{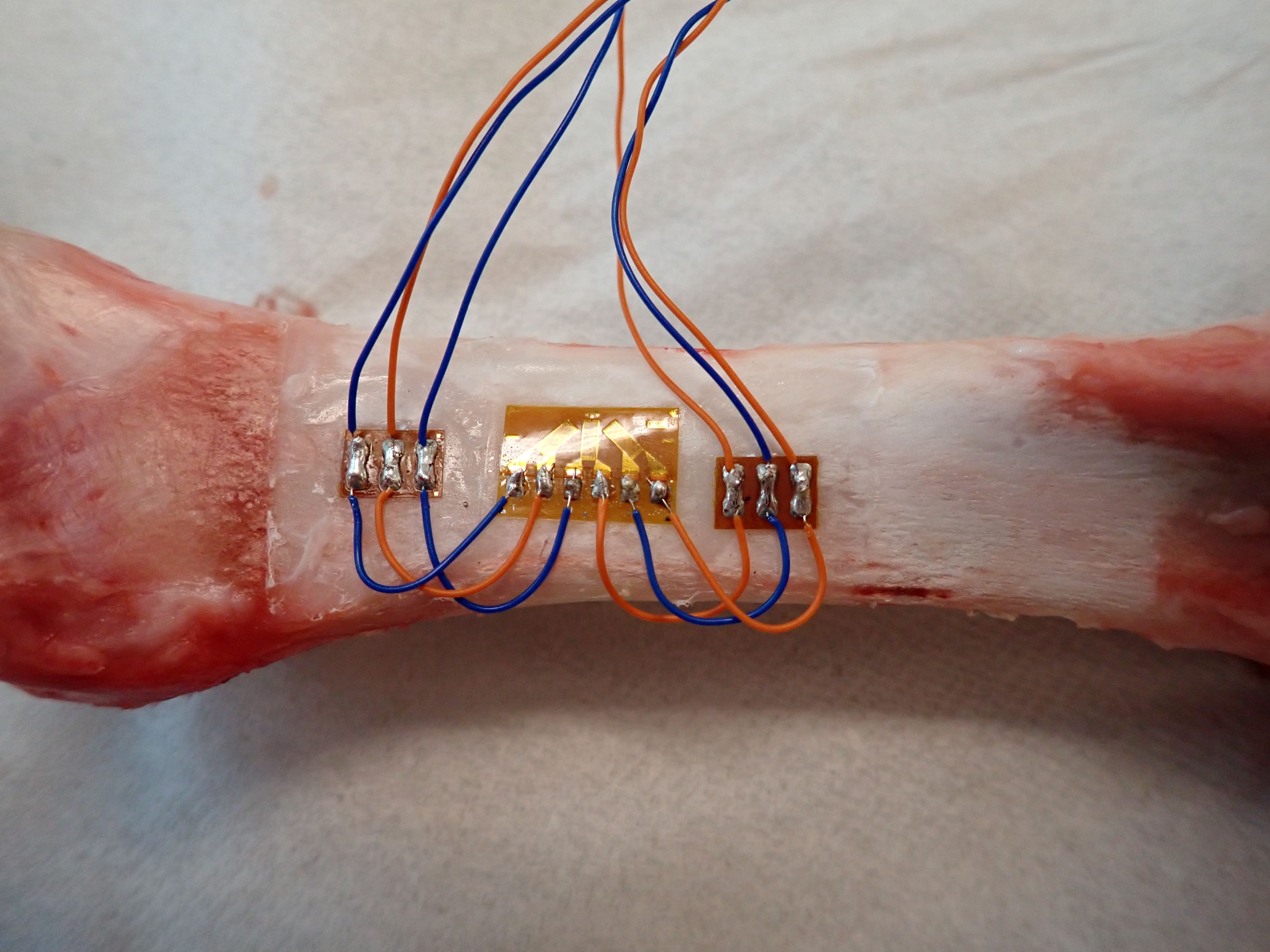}%
        \caption{Dissected tibia with a rectangular strain rosette affixed to a flat region.}
        \label{fig:tib_rosette}
    \end{subfigure}
    \caption{Strain rosettes for the fibula (a) and tibia (b) after preparing the bone surface, affixing the rosettes, soldering the terminal wires, and sealing the rosette.}%
    \label{fig:rosettes}
\end{figure}

For each fibula the total length and cross-section was measured \citep{Asabe2005}. The cross-section was determined by taking both the medial-lateral and anterior-posterior diameters at 5 locations along the elliptical cross-section and averaging the results. The diameter of an equivalent area circle was then computed and used to determine the support distance required to obtain a support length to bone diameter ratio greater than 10 as required by the ASABE standard \citep{Asabe2005}. Prior to loading, the fibula was placed on the supports and the location of the site under the center driver relative to the proximal end was marked and recorded. The center location was then prepared \citep{Zdero2017a} and a rectangular strain rosette with a resistance of 350 $\Omega$ (CEA-13-125UR-350, Intertechnology, Toronto, Ontario) was affixed with M-Bond 200 Adhesive Kit (Intertechnology, Toronto, Ontario) and sealed with M-Coat A (Intertechnology, Toronto, Ontario), as shown in \cref{fig:fib_rosette}. A flat section of the tibia was similarly prepared and affixed with a strain rosette (CEA-13-125UR-350, Intertechnology, Toronto, Ontario) for use as a temperature compensation gauge, see \cref{fig:tib_rosette} \citep{micro-measurementsStrainGageThermal2014,micro-measurementsThreeWireQuarterBridgeCircuit2015}. Each stain gauge within the rosette was connected to a National Instruments multi-channel strain data acquisition module using a two-wire quarter-bridge configuration with a temperature compensation gauge \citep{micro-measurementsThreeWireQuarterBridgeCircuit2015,Zdero2017,Zdero2017a}. The two wire configuration was chosen due to the short lead length and low lead resistance of 0.1 $\Omega$ relative to the strain gauge resistance of 350 $\Omega$.

The fibula was then placed back onto the supports and the strain rosette was aligned with the driver facing downward opposite to the drivers point of contact. An initial preload of 5 N was used and the system was paused so the distance from proximal and distal ends of the fibula to the supports could be measured safely. A compressive load was then applied in displacement control \citep{Asabe2005} until the defined displacement limit or half of the expected failure load was reached to ensure that loading could be repeated and plastic deformation did not occur. The expected failure load and displacement were calculated as per the ASABE standard using the measured cross-section for each fibula \citep{Asabe2005}. An estimated elastic modulus of 6 GPa was used for all specimens, which was determined by testing a single pilot specimen to failure \citep{Asabe2005}. From pilot testing it was also determined that an approximate tensile strength of 93 MPa gave reasonable predictions \citep{morganBoneMechanicalProperties2018}. Once the load or displacement limit was reached a constant displacement was held for 5 seconds before unloading in displacement control. Load and displacement were recorded throughout the loading cycle by the ElectroForce 3510 (TA Instruments, New Castle, Delaware) while strain data was recorded. The loading procedure was performed a total of three times for each fibula and results were averaged across trials.

\subsection{Imaging}
Prior to thawing and dissection, clinical-quality qCT scans were acquired for each of the fresh-frozen porcine hind-limbs, using the acquisition and reconstruction parameters detailed in \cref{tab:ct_scans}. All images were taken with a Model 3 QCT Phantom (Mindways, Austin, Texas) present within the field of view. CT images were then segmented using thresholding, region growing, and manual editing with Mimics (Materialise, Leuven, Belgium).

\begin{table}[!htbp]
    \centering
    \caption{Summary of CT Scan Acquisition metadata.}
    \begin{tabular}{lcc}
        \toprule
        \textbf{Parameter}                             & \textbf{Specimens} & \textbf{Value}                 \\
        \midrule
        CT Scanner                                     & All                & Toshiba Aquilion 64            \\
        Institution                                    & All                & A64S WAVES VETERINARY HOSPITAL \\
        Acquisition Mode                               & All                & Helical CT                     \\
        Beam Energy                                    & All                & 120 KVp                        \\
        Slice Thickness                                & All                & 1.0 mm                         \\
        Exposure Time                                  & All                & 500                            \\
        \multirow{3}[0]{*}{Pixel Spacing (Resolution)} & 5-11               & 0.488 $\times$ 0.488 mm        \\
                                                       & 3                  & 0.461 $\times$ 0.461 mm        \\
                                                       & 4                  & 0.406 $\times$ 0.406 mm        \\
        Total Pixels (Matrix Size)                     & All                & 512 $\times$ 512               \\
        Convolution Kernel                             & All                & FC30 (Bone)                    \\
        Tube Current                                   & All                & Auto-adjusted to specimen      \\
        \bottomrule
    \end{tabular}%
    \label{tab:ct_scans}%
\end{table}%

\subsection{Partial Volume Correction Algorithm}
The CT images and associated binary segmentation masks were loaded into Python using a custom library for DICOM processing based on pydicom \citep{masonPydicomPydicomPydicom2022} and VTK \citep{schroederVisualizationToolkitObjectoriented2006}. The set of interior voxels \textbf{Q} was defined by performing binary morphological erosion using a 3 $\times$ 3 $\times$ 3 binary kernel with a square connectivity of 1 on the segmentation mask. The set of surface voxels \textbf{S} was then defined as the set difference of the segmentation mask and \textbf{Q}. For each voxel x, a 26-connected neighbourhood $\mathbf{P}(x)$ was defined, which contains all voxels connected by a face, edge, or corner. For each voxel x in \textbf{S} a new HU value was calculated according to \cref{eqn:new_hu,eqn:current_hu,eqn:idw,eqn:idw_weights,eqn:dist} with p = 2.

\begin{equation}
    HU(x) =
    \begin{cases}
        max(h(x), u(x)) & \text{if}~ u(x) \neq 0, \\
        h(x)            & \text{if}~ u(x) = 0
    \end{cases}\label{eqn:new_hu}
\end{equation}
where
\begin{gather}
    h(x)  = \text{current HU value of voxel x}\label{eqn:current_hu}\\
    u(x) =
    \begin{cases}
        \frac{\sum w_{i}(x) \cdot HU_{i}}{\sum w_{i}(x)} & \text{if}~ d(x, x_i) \neq 0 \land \sum w_{i}(x) \neq 0, \\
        HU_{i}                                           & \text{if}~ d(x, x_i) = 0 \lor \sum w_{i}(x) = 0
    \end{cases}\label{eqn:idw}\\
    w_{i}(x) =
    \begin{cases}
        \frac{1}{d(x, x_{i})}^{p} & \text{if}~ x_i \in \mathbf{Q} \land x_{i} \in \mathbf{P}(x),     \\
        0                         & \text{if}~ x_i \notin \mathbf{Q} \lor x_{i} \notin \mathbf{P}(x)
    \end{cases}\label{eqn:idw_weights}\\
    d(x,x_{i}) = \text{distance between $x$ and $x_{i}$}\label{eqn:dist}
\end{gather}

The interpolation kernel was defined using a neighbourhood of voxels instead of a fixed radius due to the possibility of anisotropic voxel sizes in CT images with variable slice thickness. The connectivity of 26 for \textbf{P(x)} was chosen to balance minimizing discontinuities in the HU field in all 3 dimensions while still limiting the interpolation to a localized area due to the highly heterogenous nature of bone. Similarly, all voxels in \textbf{S} are assigned a weight of 0, due to the likelihood of being subject to PV effects. As a consequence of the weighting, neighbourhood size, and \cref{eqn:new_hu} any surface voxel x with no interior voxels in its neighbourhood \textbf{P(x)} will have no correction applied as u(x) will equal 0. While this does mean that voxels in thin regions, such as the blade of a scapula, will not be corrected it also means the method makes no assumption that bone material properties are smooth or continuous over volumes larger than \textbf{P(x)} as adjacent thicker regions could be anatomically different or a considerable distance away. Finally, \cref{eqn:new_hu} was formulated to prevent artificially lowering the HU value of surface voxels that are accurately segmented and not subject to PV effects by only ever increasing the HU value as bone has the highest HU value amongst natural tissues and thus PVs on bone surface voxels always decrease the voxel's intensity. After applying the PV correction method, the CT images were saved as DICOM images for use in established FE modelling workflows.


\subsection{Finite Element Modelling}
For each source CT image, a triangulated surface mesh was generated from the segmentation mask using Mimics (Materialise, Leuven, Belgium). The triangular surface meshes were then imported into SolidWorks (Dassault Systemes, Velizy-Villacoublay, France) along with the models of the rollers used as the supports and driver of the three-point bending test jig. A rigid body motion simulation was then performed to improve fidelity of the computational model regarding how each fibula lay on the support rollers. Roller to roller distance was set to the experimentally measured support span length. The fibula surface mesh was positioned with the medial or lateral surface facing the driver, in accordance with the experimental set up for each specimen. The measurements to the proximal and distal ends of the fibula were scaled to account for the difference in mesh geometry and the measured experimental geometry (bone length) and used to constrain the position of the fibula on the rollers. Finally the preload step was simulated by applying a constant displacement of 10 mm/min to the driver until contact with the fibula was achieved. The resulting configuration of the fibula, support rollers, and driver for each specimen was then exported for use in creating the finite element meshes. 

An initial n-points registration and subsequent iterative closest point registration was used to align the assembly containing the fibula, driver, and support rollers for each specimen with the source CT image. As the fibula surface mesh was unchanged a registration error of 0 was achieved for each specimen. The triangulated surface mesh of the fibula was then imported into 3-Matic (Materialise, Leuven, Belgium), where it was remeshed to achieve a uniform edge length before generating a quadratic tetrahedral mesh. As the PV corrected and the source CT images share the same voxel architecture and segmentation masks, the quadratic tetrahedral mesh generated for each source CT image was also used for the corresponding PV corrected images.

Material properties from both the source and PV corrected CT images were then assigned to the mesh using Mimics (Materialise, Leuven, Belgium), which implements an element-based MMM. HU values were first calibrated and converted to equivalent density according to the MindWays phantom-derived CT calibration curve \citep{mindwayssoftwareinc.BoneMineralDensitometry2011}.

As the specimens were of varying skeletal maturity and there are no well-validated density-modulus relationships for porcine fibulae in the literature, we computationally derived specimen-specific density-to-modulus relationships according to the response surface methodology published by Eberle et al (2013) \citep{eberleIndividualDensityElasticity2013}. The response surface aimed to minimize the difference between the FE computed and experimentally measured root-mean-square-error (RMSE) of the maximum principal strain at the strain gauge location and the whole bone stiffness. Furthermore, we adapted Eberle et al.'s method by sampling a greater number of points (i.e. xx points) in order to calculate our response surface. As well, we incorporated an inequality constraint to ensure the maximum elastic modulus for each specimen was not greater than 20 GPa, which was the maximum value reported it the literature for porcine bone \citep{fengMechanicalPropertiesPorcine2012}. Finally, we reduced the lower bound for the power law coefficient A (from $A(rho)^b)$ to 3000 MPa instead of 5000 MPa based on literature for tibia relationships \citep{grantComparisonMechanicalUltrasound2014,snyderEstimationMechanicalProperties1991}. Response surface design and optimization was done using R and NLopt \citep{lenthResponsesurfaceMethodsUsing2009,NLopt,DIRECT}.

As discussed in the introduction, work by Pakdel and Helgason has previously shown that PVC improved results and more accurately reflect reality. Thus, in order to isolate the effect that the PVC method we have developed has and avoid other confounding factors, only one density-to-modulus relationship was derived for each specimen and it was based exclusively on the PV corrected images. Specifically, the response surface method was only applied to simulations derived from PV corrected images but the resulting 'optimal' density-to-modulus relationships were used for material property assigned for FE simulations derived from both the PV corrected and the uncorrected (i.e. raw) CT images.

After determination of the specimen-specific density-to-modulus relationship for each specimen, material properties were assigned to each mesh. To minimize averaging effects during the element-based material assignments that could reduce the effectiveness of the PVC algorithm, elastic moduli were grouped into 100 bins for trabecular bone and 100 bins for cortical bone. Finally, a Poisson's ratio of 0.3 was set for all material groups \citep{Studders2020a} .

Once material properties were assigned, FE model assemblies were import into Abaqus (Dassault Systemes, Velizy-Villacoublay, France) to apply boundary conditions and solve. All of the rollers were modelled as discrete rigid elements as the stiffness of 304 stainless steel greatly exceeded that of the fibula. Each supporting roller was constrained to have zero displacement in all degrees of freedom except for rotation about the long axis of the roller, while the driver was constrained to have zero displacement in all degrees of freedom except along the line of action of the MTS mover, and the fibula was unconstrained. Contact pairs between the fibula and each of the supports as well as the fibula and the driver were created and assigned the coefficient of friction $\mu = 0.37$ \citep{lopez-camposFiniteElementStudy2018} using a tangential behavior. A load with magnitude equal to the average peak experimental load was then applied along the driver's line of action. The average peak load was used because, during the experimental dwell phase (i.e. after the peak was reached) the displacement and strain held relatively constant but the load showed a gradual decline due to bone stress relaxation.

Although the bending tests were performed in displacement control applying a load in the FE simulations was chosen because the dependant experimental variable was principal strain, which would be constant regardless whether the model is derived from raw or PV corrected images if a constant displacement was used. Automatic stabilization in Abaqus was enabled to apply damping during the initial solving step to improve stability when resolving contact. The damping was gradually reduced over the first solver step such that it was no longer applied in subsequent solver steps. It was determined that the models could be solved without damping but converged to a solution much slower, while still reaching the same result as with damping.

After solving, strain data was extracted from mesh elements in the location and shape of the strain rosettes, and the same elements were selected for the original and PV corrected simulations. The maximum principal strain was calculated as the mean maximum principal strain of the elements in the selected area.

\begin{figure}[!htbp]
    \centering
    \includegraphics[width=\textwidth]{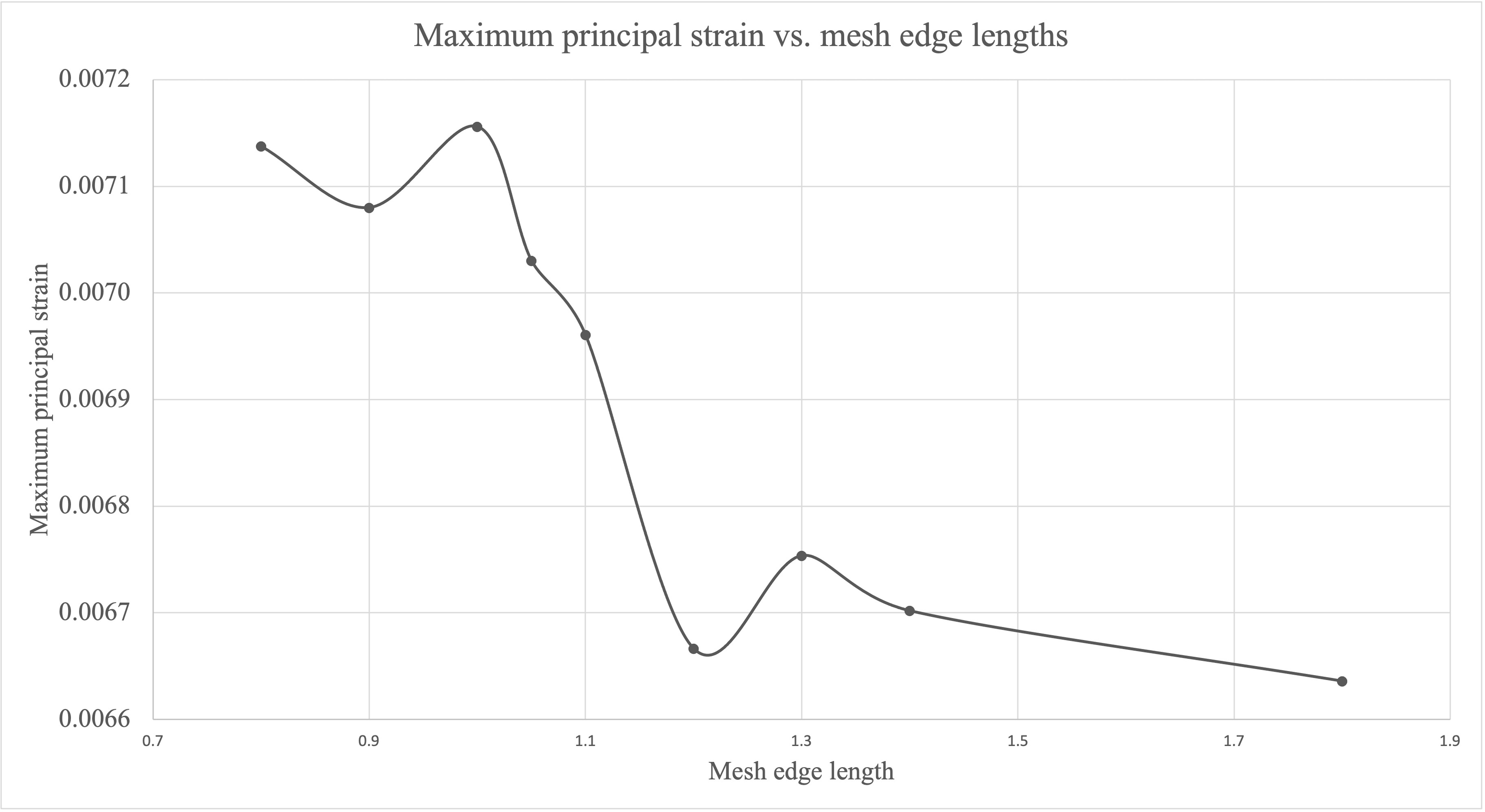}
    \caption{Maximum principal strain from different mesh edge lengths.}%
    \label{fig:MCAGraph}
\end{figure}

Mesh convergence was investigated using an FE simulation derived from the original CT images for specimen 7 and plotting the mean maximum principal strain against the mean element edge length, see \cref{fig:MCAGraph}. The minor oscillation in \cref{fig:MCAGraph} can be attributed to the variable topology of each mesh making it impossible to select the same elements to compute the mean maximum principal strain for each mesh. Similarly, selecting elements that had an equal area to the strain rosette was not always possible for meshes with a larger average element edge length. Given the trend in \cref{fig:3bend_jig}, a target maximum edge length of 1.0 mm was chosen to balance computational cost and accuracy. This result also agreed with previous findings in the literature that FE meshes achieve good convergence when the average element edge length is approximately equal to the CT image slice thickness \citep{perillo-marconeAssessmentEffectMesh2003}.

\subsection{Statistical Comparison Between Experimental Testing and Finite Element Analysis}

The experimental mean maximum principal strain was calculated over 100 data points from the time the peak load for each specimen was reached in each of the three loading tests. An average of the experimental mean maximum principal strain over the three loading tests was then calculated and used for comparison with the computational strain results. The relative error between experimental and computational strain was then calculated. After checking normality using the Shapiro-Wilks test and confirming the assumption of equal variance with Levene's test, a single tailed paired t-test with $h_{0}: d = 0$ and $h_{1}: d > 0$ where $d = \epsilon_{Source} - \epsilon_{PVC}$ and $\alpha = 0.05$ was performed to determine if FE simulations derived from PV corrected CT images had lower relative error than FE simulations derived from the raw CT images. Analysis was performed using the scipy.stats and researchpy packages in Python \citep{2020SciPyNMeth,bryantResearchpyResearchpyProduces2018}.

\section{Results}

\subsection{Specimen-Specific Density-to-Modulus Relationships}
It was found that across the eight specimen the optimized specimen-specific density-to-modulus relationships calculated using the response surface method had highly variable constants, which agreed with qualitative observations indicating that the skeletal maturity of the specimen was highly variable (Table~\ref{tab:rho_modulus}).

\begin{table}[!htbp]
    \centering
    \caption{Summary of computationally derived specimen-specific density-modulus relationship coefficients.}
    \begin{tabular}{ccc}
        \toprule
        \multicolumn{1}{c}{\textbf{Specimen}} & \textbf{A (MPa)} & \textbf{B} \\
        \toprule
        3                                     & 12277.42         & 0.994193   \\
        4                                     & 13684.27         & 0.88775    \\
        5                                     & 11114.34         & 1.295186   \\
        6                                     & 10306.96         & 1.441808   \\
        8                                     & 12756.7          & 1.080887   \\
        9                                     & 12761.89         & 1.091541   \\
        10                                    & 10975.18         & 1.461723   \\
        11                                    & 9010.101         & 1.748485   \\
        \bottomrule
    \end{tabular}%
    \label{tab:rho_modulus}%
\end{table}%

\subsection{Strain}
\begin{figure}[!htbp]
    \centering
    \includegraphics[width=14cm]{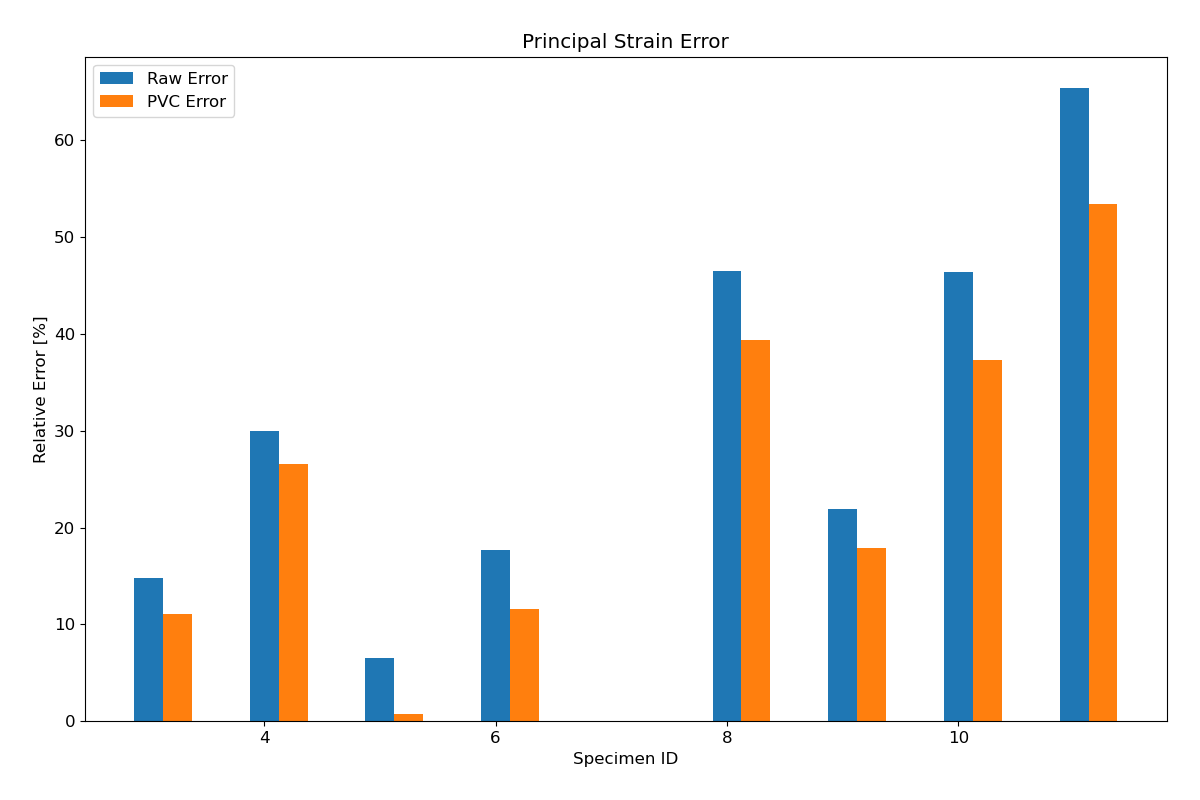}
    \caption{Relative difference between simulated and experimental maximum principal strain.}
    \label{fig:strain_err}
\end{figure}

The maximum principal strains from the experimental testing, the original image derived simulations, and the PVC image derived simulations are tabulated in Table~\ref{tab:strain_data}, along with the relative differences between the simulated and experimental data. The relative differences between the two methods are graphically shown in Figure~\ref{fig:strain_err}. It can be seen that the strain error compared to the experimental data is lower for all specimens when using the PVC image derived vs original image derived models with the difference between the models averaging 6\% (range: 3 - 12\%). This difference in relative strain error was found to be statistically significant (p$<$0.05, see Table~\ref{tab:strain_t_test}). Descriptive statistics regarding data normality and equality of variance for all data can be see in Table~\ref{tab:desc_stats}.


\begin{table}[!htbp]
    \centering
    \caption{Maximum principal strain from experimental testing, original simulation, and PVC simulation, and the relative differences between the simulated and experimental data.}%
    \begin{tabular}{cccccc}
        \toprule
        \textbf{Specimen} & \multicolumn{3}{c}{\textbf{Max principal strain}} & \multicolumn{2}{c}{\textbf{Relative Difference}}                                                                    \\
        \textbf{}         & \multicolumn{1}{c}{\textbf{Experimental}}         & \multicolumn{1}{c}{\textbf{Raw}}                 & \textbf{PVC}                       & \textbf{Raw} & \textbf{PVC} \\
        \toprule
        3                 & 0.002303                                          & 0.00264308                                       & 0.002559104                        & 15\%         & 11\%         \\
        4                 & 0.002225                                          & 0.002892817                                      & 0.002816715                        & 30\%         & 27\%         \\
        5                 & 0.003236                                          & 0.003446681                                      & 0.003261663                        & 6\%          & 1\%          \\
        6                 & 0.002425                                          & 0.002853342                                      & 0.002705169                        & 18\%         & 12\%         \\
        8                 & 0.00207                                           & 0.003031263                                      & 0.002884193                        & 46\%         & 39\%         \\
        9                 & 0.001986                                          & 0.002421083                                      & 0.002341355                        & 22\%         & 18\%         \\
        10                & 0.00189                                           & 0.002766969                                      & 0.002595284                        & 46\%         & 37\%         \\
        11                & 0.001876                                          & 0.003101739                                      & 0.002877716                        & 65\%         & 53\%         \\
        \midrule
                          &                                                   &                                                  & \multicolumn{1}{r}{\textbf{Mean:}} & 31.13\%      & 24.75\%      \\
        \bottomrule
    \end{tabular}
    \label{tab:strain_data}
\end{table}


\begin{table}[htbp]
    \centering
    \caption{Paired T-Test Results for Strain Error}
    \adjustbox{max width=\textwidth}{%
        \begin{tabular}{ccccccc}
                                & \textbf{N} & \textbf{Mean}                                   & \textbf{Variance}                            & \textbf{SD}                                  & \textbf{SE} & \textbf{95\% Conf. Interval} \\
            \midrule
            \textbf{PVC}        & 8          & 0.247475                                        & 0.031085                                     & 0.176309                                     & 0.062334    & [0.1001,0.3949]              \\
            \textbf{Raw}        & 8          & 0.311274                                        & 0.039712                                     & 0.199278                                     & 0.070456    & [0.1447,0.4779]              \\
            \textbf{PVC - Raw}  & -          & -0.063798                                       &                                              & 0.029546                                     & 0.010446    & [-0.0885,-0.0391]            \\
            \midrule
                                &            &                                                 &                                              &                                              &             &                              \\
            \midrule
            \textbf{Difference} &            & \multicolumn{1}{c}{\textbf{PVC - Raw $\neq$ 0}} & \multicolumn{1}{l}{\textbf{PVC - Raw $<$ 0}} & \multicolumn{1}{l}{\textbf{PVC - Raw $>$ 0}} &             &                              \\
            \textbf{P-Value}    &            & 0.000488                                        & 0.000244                                     & 0.999756                                     &             &                              \\
            \bottomrule
        \end{tabular}%
    }%
    \label{tab:strain_t_test}%
\end{table}%

\begin{table}[htbp]
    \centering
    \caption{Descriptive statistics of the sets of relative differences.}
    \begin{tabular}{ccccccc}
                                & \multicolumn{2}{c}{\textit{\textbf{Strain Error}}} & \multicolumn{2}{c}{\textit{\textbf{Modulus Error}}} & \multicolumn{2}{c}{\textit{\textbf{RMSE}}}                                              \\
        \cmidrule{2-7}          & \textbf{Raw}                                       & \textbf{PVC}                                        & \textbf{Raw}                               & \textbf{PVC} & \textbf{Raw} & \textbf{PVC} \\
        \midrule
        \textbf{Shapiro-Wilks } & 0.6075                                             & 0.8324                                              & 0.8976                                     & 0.8652       & 0.5273       & 0.8265       \\
        \textbf{Levene's Test}  & \multicolumn{2}{c}{0.7044}                         & \multicolumn{2}{c}{0.757}                           & \multicolumn{2}{c}{0.9456}                                                              \\
        \bottomrule
    \end{tabular}%
    \label{tab:desc_stats}%
\end{table}%

\begin{figure}[!htbp]
    \centering
    \includegraphics[width=14cm]{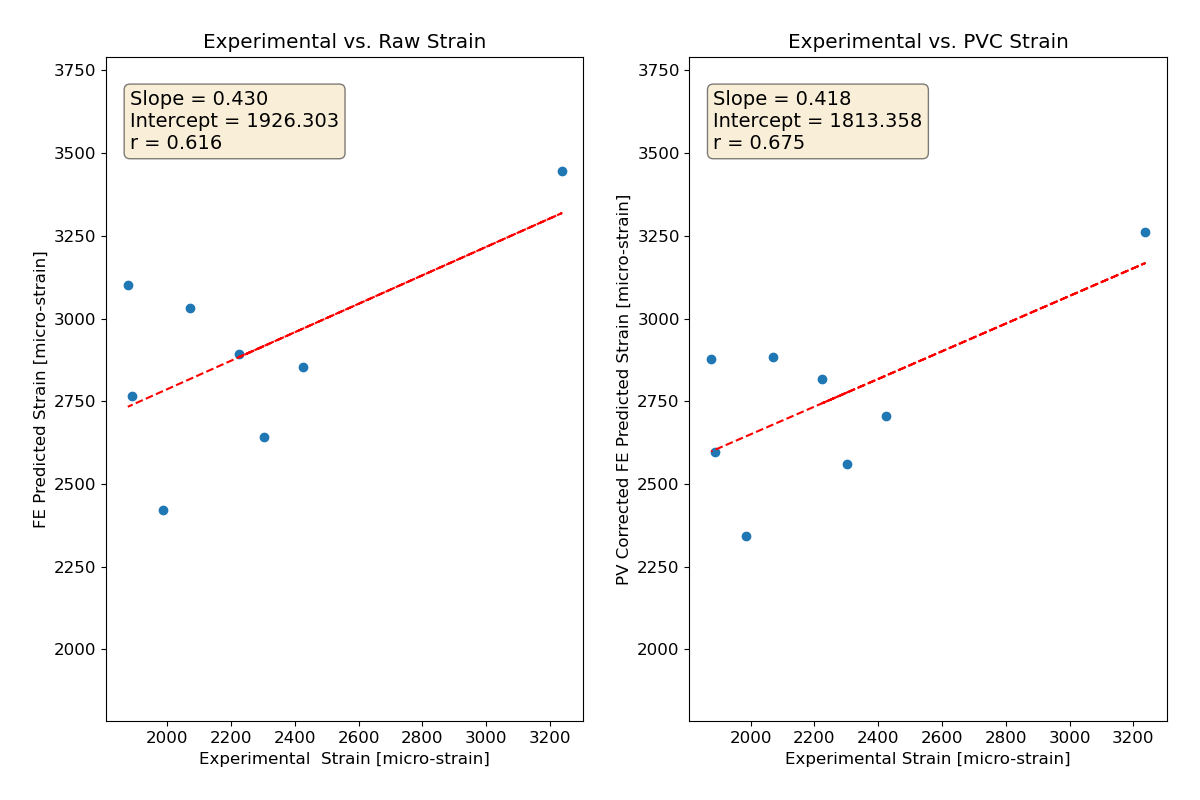}
    \caption{Linear correlation of experimental strain compared to computational determined strain from the raw and partial volume corrected simulations.}
    \label{fig:strain_correlation}
\end{figure}

Figure~\ref{fig:strain_correlation} illustrates that the strain from the PVC models produces a slightly more linear relationship to the experimental strain (r=0.675) compared to the strain from the original models (r=0.616). However, the linear fit for both types of model has a slope of $\approx$0.4 when it would be expected to be 1.

\subsection{Modulus}
\begin{figure}[!htbp]
    \centering
    \includegraphics[width=14cm]{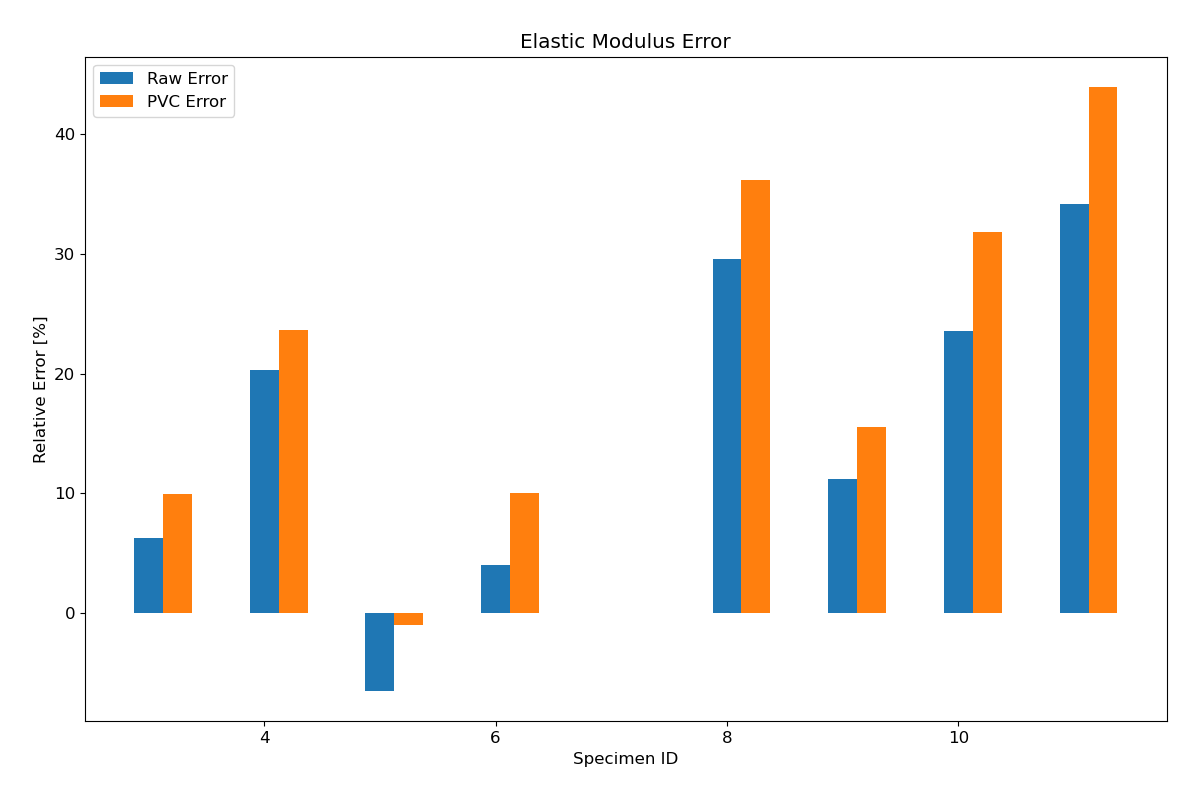}
    \caption{Relative difference between simulated and experimental whole bone elastic modulus.}
    \label{fig:modulus_err}
\end{figure}

In addition to comparing the models' ability to replicate the maximum principal strain, the whole bone moduli of each specimen were compared between the experimental data and the PVC-derived and original image-derived models. In this case, the modulus calculated from models using the original images more accurately matched the experimental data with an average percent error of 15.31\% while the PVC-derived models had 6\% higher error (Figure ~\ref{fig:modulus_err}). This difference in relative modulus error was found to be statistically significant (p$<$0.05, see Table~\ref{tab:modulus_t_test}).

\begin{table}[htbp]
    \centering
    \caption{Paired T-Test Results for Modulus Error}
    \adjustbox{max width=\textwidth}{%
        \begin{tabular}{ccccccc}
                                & \textbf{N} & \textbf{Mean}                                   & \textbf{Variance}                            & \textbf{SD}                                  & \textbf{SE} & \textbf{95\% Conf. Interval} \\
            \midrule
            \textbf{PVC}        & 8          & 0.212662                                        & 0.023474                                     & 0.153211                                     & 0.054168    & [0.0846, 0.3407]             \\
            \textbf{Raw}        & 8          & 0.153135                                        & 0.019412                                     & 0.139328                                     & 0.04926     & [0.0367, 0.2696]             \\
            \textbf{PVC - Raw}  & -          & 0.059526                                        &                                              & 0.02235                                      & 0.007902    & [0.0408, 0.0782]             \\
            \midrule
                                &            &                                                 &                                              &                                              &             &                              \\
            \midrule
            \textbf{Difference} &            & \multicolumn{1}{c}{\textbf{PVC - Raw $\neq$ 0}} & \multicolumn{1}{l}{\textbf{PVC - Raw $<$ 0}} & \multicolumn{1}{l}{\textbf{PVC - Raw $>$ 0}} &             &                              \\
            \textbf{P-Value}    &            & 0.000134                                        & 0.999933                                     & 0.000067                                     &             &                              \\
            \bottomrule
        \end{tabular}%
    }%
    \label{tab:modulus_t_test}%
\end{table}%

\begin{figure}[!htbp]
    \centering
    \includegraphics[width=14cm]{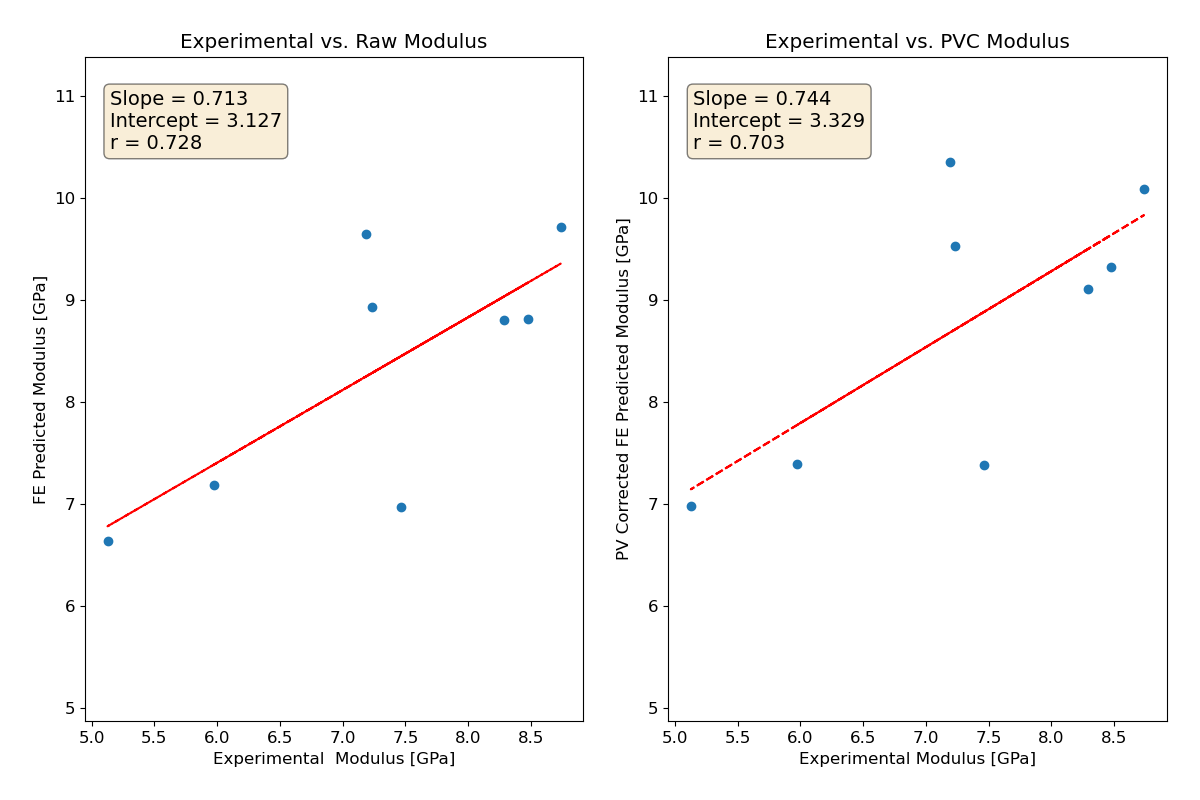}
    \caption{Linear correlation of experimental elastic modulus compared to computaional determined modulus from the raw and partial volume corrected simulations.}%
    \label{fig:modulus_correlation}
\end{figure}

Figure~\ref{fig:modulus_correlation} illustrates that the modulus from the original image-derived models produces a slightly more linear relationship to the experimental modulus (r=0.728) compared to the modulus from the PVC models (r=0.703). However, the linear fit for both types of model has a slope of $\approx$0.7 when it would be expected to be 1.

\subsection{RMSE}
\begin{figure}[!htbp]
    \centering
    \includegraphics[width=14cm]{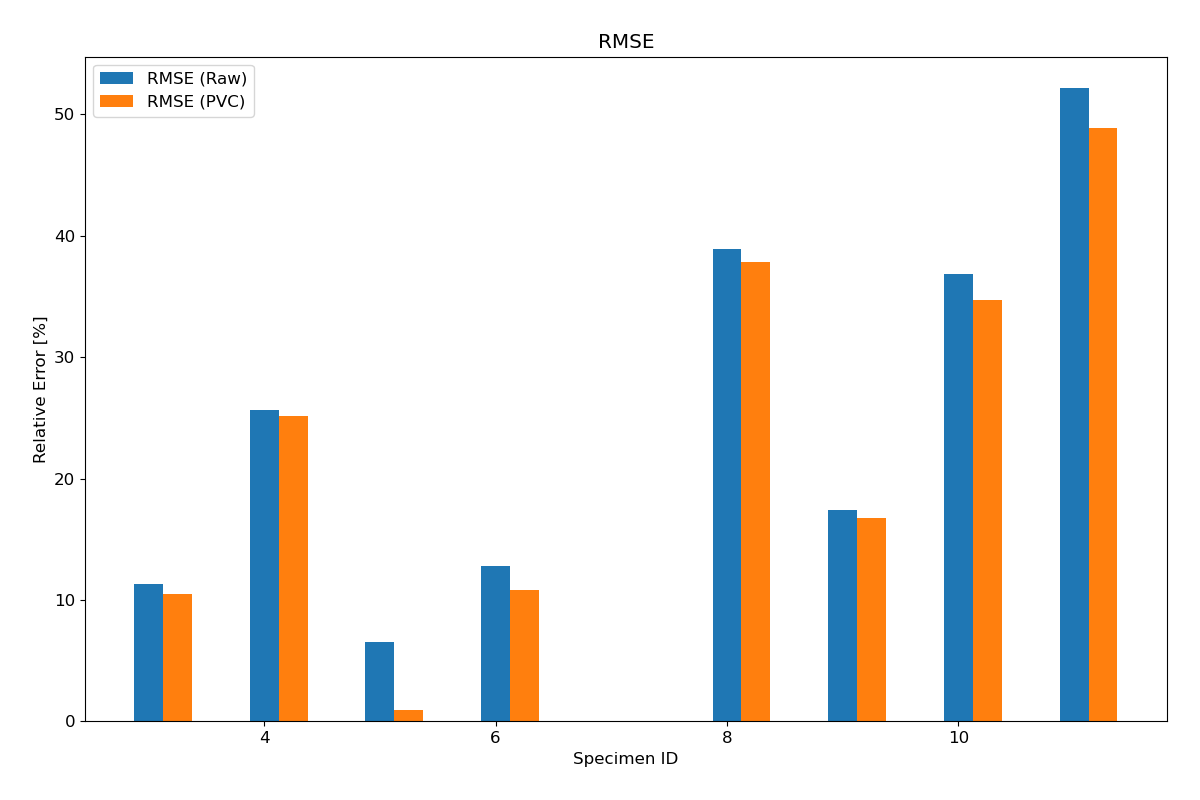}
    \caption{Relative difference between simulated and experimental root-mean-square-error as a combined measure of strain and modulus error.}%
    \label{fig:rmse_err}
\end{figure}

Assessing the RMSE percent relative error that combines the relative strain and modulus errors (Figure ~\ref{fig:rmse_err}) it can be seen that the PVC-derived models produce slightly lower error compared to the experimetal results, averaging 23.20\%. The models derived from the original images produced RMSE that was 2\% higher than that found with the PVC-derived models. This difference in RMSE was found to be statistically significant (p$<$0.05, see Table~\ref{tab:rmse_t_test}).

\begin{table}[htbp]
    \centering
    \caption{Paired T-Test Results for RMSE}
    \adjustbox{max width=\textwidth}{%
        \centering
        \begin{tabular}{ccccccc}
                                & \multicolumn{1}{c}{\textbf{N}} & \multicolumn{1}{c}{\textbf{Mean}}               & \multicolumn{1}{c}{\textbf{Variance}}        & \multicolumn{1}{c}{\textbf{SD}}              & \multicolumn{1}{c}{\textbf{SE}} & \multicolumn{1}{c}{\textbf{95\% Conf. Interval}} \\
            \midrule
            \textbf{PVC}        & \multicolumn{1}{c}{8}          & \multicolumn{1}{c}{0.232011}                    & \multicolumn{1}{c}{0.0266}                   & \multicolumn{1}{c}{0.163094}                 & \multicolumn{1}{c}{0.057662}    & \multicolumn{1}{c}{[0.0957, 0.3684]}             \\
            \textbf{Raw}        & \multicolumn{1}{c}{8}          & \multicolumn{1}{c}{0.251944}                    & \multicolumn{1}{c}{0.025785}                 & \multicolumn{1}{c}{0.160577}                 & \multicolumn{1}{c}{0.056773}    & \multicolumn{1}{c}{[0.1177, 0.3862]}             \\
            \textbf{PVC - Raw}  & \multicolumn{1}{c}{-}          & \multicolumn{1}{c}{-0.019933}                   & \multicolumn{1}{c}{-}                        & \multicolumn{1}{c}{0.01732}                  & \multicolumn{1}{c}{0.006124}    & \multicolumn{1}{c}{[-0.0344, -0.0055]}           \\
            \midrule
                                &                                &                                                 &                                              &                                              &                                 &                                                  \\
            \midrule
            \textbf{Difference} &                                & \multicolumn{1}{c}{\textbf{PVC - Raw $\neq$ 0}} & \multicolumn{1}{l}{\textbf{PVC - Raw $<$ 0}} & \multicolumn{1}{l}{\textbf{PVC - Raw $>$ 0}} &                                 &                                                  \\
            \textbf{P-Value}    &                                & 0.013957                                        & 0.006979                                     & 0.993021                                     &                                 &                                                  \\
            \bottomrule
        \end{tabular}%
    }%
    \label{tab:rmse_t_test}%
\end{table}%


\section{Discussion}

This work developed a flexible and easy to use partial volume correction algorithm to overcome blurring and PV effects at cortical bone boundaries. The algorithm works directly on the CT images themselves and yields corrected CT images, rather than making the correction during 3D mesh material assignment. This image based approach means that the algorithm can be applied for a range of application beyond just FE modeling including deep learning, and medical image recognition. In the context of  its use in FE modeling, which was our main focus, yielding corrected CT images means that this algorithm can be used with all existing CT-derived FE creation workflows and material mapping strategies. Furthermore, because the algorithm does not make assumptions about morphology, it has the potential for use across a range of bones, which was a limitation of previously published works. Specifically, the algorithm will only ever increase the density of surface voxels (i.e. suspected cortical bone) and thus, it will never make poorly segmented cortical bone even worse. For instance, in some cases a poorly segmented image will identify soft tissue as surface cortical bone and in previous methods (e.g. Pakdel et al.) the presence of the soft tissue would cause the interpolated value for neighbouring true cortical bone voxel's to decrease, which exacerbates the issue of partial volume effects. Finally, a tangential result of this work was the further validation of previously described methods for generating computationally derived specimen-specific density-modulus relationships and demonstration that additional constraints such as a maximum elastic modulus may be incorporated to yield more realistic results.

Qualitative review of the corrected images clearly demonstrates that the developed PVC algorithm was successful at correcting the PV effects and in fact had a consistent corrective effect across all specimen. For both strain and whole bone modulus results the level of error produced by either type of model was highly variable ranging from as little as 1\% up to 65\%. This large range can be attributed to the difficulty in extracting model data that exactly matches the experimental setup and due to the highly variable skeletal maturity that was qualitatively observed and confirmed by the large range in density-to-modulus relationships calculated (see Table~\ref{tab:rho_modulus})

Partial volume correction resulted in improved surface strain values that more closely matched the experimental results with a 6.4\% improvement compared to the results produced with the uncorrected, original models. This level of improvement in strain is similar to the mean improvement of 8\% reported by Helgason et al. (2008) although they reported a mean increase in modulus error of 10\% which was greater than the mean increase in modulus error of 6\% observed with our method. Despite this improvement, the strain calculated using the PVC models still averaged 25\%, which is attributable to the difficulty in extracting strain values from the FE models that exact match the experimental strain gauges geometry and location, as well as the effect of needing to create a custom specimen-specific density-to-modulus relationship rather than being able to rely on previously validated relationships. Despite these complications the improvement in strain error from the PVC models was found to be statistically significant.

In addition to the discrete strain error values for each specimen, it is also useful to assess how the computational and experimental strains correlate across all specimen. In this analysis we found that neither the original or PV corrected model strains achieved the desired 1-to-1 relationship with experimental strain and only produced a moderately strong correlation. These findings can be attributed to the difficulty in extracting strain values from the FE models that precisely match the experimental strain gauge location which can cause the extracted value to be higher or lower than it would be if precisely matched which increases the randomness of the value thus affecting the linear fit and correlation. As well, the highly variable skeletal maturity of the specimen resulting in very different density-to-modulus relationships (see Table~\ref{tab:rho_modulus}) would also effect these results.

Application of partial volume correction also resulted in increases in the calculated whole bone modulus of each specimen as one would expect when the surface cortical bone layer HU values are increased. However, the specimen-specific density-to-modulus relationships resulted in consistent over-estimates of the whole bone modulus for all specimen even when using the uncorrected, original images. As a result, the PVC derived FE models actually had higher error compared to the uncorrected models with an average of 6\%. This unexpected negative effect of the PV correction is most likely caused by the response surface method used to derived specimen-specific density-to-modulus relationships, which tried to minimize both strain error and whole bone modulus, which change in opposing ways. For instance, as surface modulus increases whole bone modulus increases but surface strain decreases.

Considering the linear fit and correlational findings for whole bone modulus, it can be seen that both types of models produce results that are closer to the expected pattern as compared to the strain results but there was minimal difference between the two types of models. Specifically, the linear relationship had a slope of approximately 0.7, which more closely matched our expectation of a 1-to-1 relationship than we observed for strain results. As well, we observed correlations of approximately 0.7 which is nearing a strong relationship. It is believed that these results would be even better if it wasn't for the same complicating factors noted above for the strain related findings.

Considering both the strain and modulus errors together through the RMSE, it can be seen that the PV corrected models did produce a 2\% improvement over the results for the original models compared to the experimental results. This overall improvement is minimized by the overestimate of whole bone modulus that the specimen-specific density-to-modulus relationships produced.


\textbf{Impact of the Work}
The developed algorithm is expected to produce a number of impacts due to its improvements and differences in approach compared to previously published methods. First, this code is standalone and works on the images themselves, which allows for the method to easily be integrated into existing workflows regardless of the preferred software for segmentation, meshing, and material assignment. This will provide greater user freedom to conduct their research compared to previously published PVC methods (Helgason, Pakdel) that were integrated with specific material mapping strategies or packaged as part of dedicated software (MITK-GEM), thus requiring significant changes to existing workflows. Second, the described method requires less user input than in previously published methods such as Pakdel et al.'s that required deconvolution prior to using the method, which in turn required the user to have a significant degree of expertise/knowledge in order to optimize the result. The present method demonstrated that it is possible to perform a partial volume correction directly on the CT data while improving the accuracy of the results with clinical grade CT scans that have not undergone deblurring or deconvolution as recommended by Pakdel prior to applying their PVC method. Third, whereas previous methods have either not been open-source or difficult to modify due to their integration with other packages, this method will be released in a GitHub repo that provides the code for performing the PVC on any CT images formatted as a DICOM stack (\url{https://github.com/adbeagley/pvcpy}). The code will be released under an open-source license that allows other researchers to adopt it and/or iterate on the methods.

\textbf{Limitations}
As with any computational method, the algorithm developed here does have limitations. First, in the case of a poor segmentation which has poorly identified the bone surface such that the surface voxels and their adjacent interior voxels are actually located in regions of soft tissue, the method will be unable to accurately correct the HU intensity as the interior voxels do not represent cortical bone that has been less affected by partial volume artifacts. Second, this work has only characterized the effect of the partial volume correction algorithm with a single type of mesh (i.e. 10 node tetrahedral volume elements) and material mapping strategy (i.e. the element-based method used by Mimics). Previous literature has shown that the element type and material mapping strategy does influence the effectiveness of PVC; however, those methods applied the correction to the meshes and thus it is expected that meshing choices would effect results. Conversely, this PVC method works on the images and thus although later meshing choices may effect results it should not be effected by the PVC method. Third, there is currently no well validated density-to-modulus relationship for porcine bone, especially in circumstances of variable skeletal maturity as observed in this study. Therefore, computationally derived specimen-specific density-to-modulus relationships were determined based on the PVC images rather than using an empirical relationship. Use of an empirical relationship would have been preferable as it would have allowed us to characterize the PVC algorithm's ability to match true bone density rather than being limited to characterizing the algorithm's relative improvement compared to the uncorrected images.

\textbf{Future Work}
To further validate this algorithm, future work will focus on assessing the effect of pairing the PV correction with multiple material assignment strategies that including both element and node based methods. This will provide greater confidence that this new method is in fact agnostic to the material assignment method used. As well, in these future works, the method will be applied to human bones with well-validated empirical density-to-modulus relationships that will enable a direct assessment of the method's ability to correct surface voxel density to the true values.

\textbf{Conclusion}
This work has developed and preliminarily validated a partial volume correction algorithm that works directly on CT images, is easy to use, and can be integrated with any existing FE workflows. As well the method can be used in other applications that work with CT images and would benefit from the removal of PV effects including deep learning based semantic segmentation.

\bibliography{references/pvcRefs.bib}  


%
\end{document}